\documentclass[12pt]{article}

\usepackage{setspace}
\usepackage{amsmath,amssymb,amscd}
\usepackage{amsfonts}
\usepackage{color}

\usepackage{graphicx}

\let\OLDthebibliography\thebibliography
\renewcommand\thebibliography[1]{
  \OLDthebibliography{#1}
  \setlength{\parskip}{0pt}
  \setlength{\itemsep}{2pt}
}

\usepackage[margin=1.in]{geometry}

 \newcommand{\bea}{\begin{eqnarray}}
\newcommand{\eea}{\end{eqnarray}}
\newcommand{\be}{\begin{equation}}
\newcommand{\ee}{\end{equation}}
\newcommand{\ba}{\begin{align}}
\newcommand{\ea}{\end{align}}

\newcommand{\RR}{\mathbb{R}} % Reali
\newcommand{\ZZ}{\mathbb{Z}} % Interi
\newcommand{\M}{\mathcal{M}}

\newcommand{\A}{\mathcal{A}}

\newcommand{\cL}{\mathcal{L}}

\newcommand{\Z}{\mathbb{Z}}

\newcommand\rref[1]{(\ref{#1})}

\usepackage{titlesec}
\titleformat*{\section}{\Large \bfseries}
\titleformat*{\subsection}{\large \bfseries}
\titleformat*{\subsubsection}{\large\bfseries}
\begin{document}

\title{\bf{Geometric Microstates for the Three Dimensional Black Hole?}}

\author{Alexander Maloney\footnote{\texttt{maloney@physics.mcgill.ca}}}
\date{}

\maketitle

\begin{center}

{\it  Department of Physics, McGill University, Montr\'eal, Canada } \\  \smallskip

%\vspace{2em} 

\end{center}
\vspace{3em}

\begin{center}
{\bf Abstract}
\end{center}

We study microstates of the three dimensional black hole obtained by quantizing topologically nontrivial geometries behind the event horizon.  In chiral gravity these states are found by quantizing the moduli space of bordered Riemann surfaces.  In the semi-classical limit these microstates can be counted using intersection theory on the moduli space of punctured Riemann surfaces.  We make a conjecture (supported by numerics) for the asymptotic behaviour of the relevant intersection numbers.  The result is that the geometric microstates with fixed topology have an entropy which grows too slowly to account for the semiclassical Bekenstein-Hawking entropy.  
%The sum over topologies leads to a divergence, which must be resolved if black hole microstates are to be described in terms of quantized geometry.
 The sum over topologies, however, leads to a divergence.  
 We conclude with some speculations about how this might be resolved to give an entropy proportional to horizon area.

\newpage

\tableofcontents

\section{Introduction}

In most quantum mechanical descriptions of black holes the geometric structure of the individual quantum microstates is unclear.  In string theory, the microstates of certain supersymmetric black holes can be described as D-Brane bound states, but although such states can be counted \cite{Strominger:1996sh}  their dynamics is difficult to study at finite coupling.  Similarly, in the AdS/CFT correspondence \cite{Maldacena:1997re} black holes are dual to quantum states in the dual CFT, but this does not give a clear description of the black hole interior.  We would like to find a black hole whose microstates can be characterized directly in geometric terms.\footnote{One notable attempt is the fuzzball proposal \cite{Mathur:2005zp}, which is somewhat similar to our approach.} This might allow us to understand, for example, the origin of black hole entropy and the emergence of classical space-time from the coarse-graining of quantum microstates.   

This note describes a somewhat speculative attempt to construct the quantum states of black holes in certain simple theories of gravity.  
We will consider theories whose degrees of freedom are purely geometric, and do not include additional matter fields such as those that arise in string theory.   We will focus on theories of gravity in three space-time dimensions which appear to possess no local degrees of freedom, making the theories relatively easy to study.

We will begin by considering a simple class of geometries which are interpreted as describing the microstates of the three dimensional black hole of Banados-Teitelboim-Zanelli (BTZ)  \cite{Banados:1992wn}.  These geometries have regular black hole horizons, but instead of a second asymptotic region they have a finite geometry with non-trivial topology hiding behind the horizon.  
The phase space of these solutions is finite dimensional, which reflects the fact that although these theories have no local degrees of freedom they still possess a finite number of ``global" degrees of freedom associated with the topology of space-time.
We will relate this phase space to the moduli space of Riemann surfaces and approximate the number of states at large genus using some known facts (and two conjectures) about the the topology of these moduli spaces.  Along the way we will make various assumptions, which we attempt to state clearly.

The result of our computation is that, when one quantizes the phase space of geometries of fixed topology, one obtains a finite dimensional Hilbert space with a discrete spectrum.  However, the number of states is too small to account for the semi-classical Bekenstein-Hawking entropy.  Moreover, the large genus limit -- discussed at the end of this paper -- remains mysterious and must be understood if we are to obtain a finite black hole entropy.  We do, however, identify a suggestive set of contributions to the number of states that give an entropy linear in the horizon area at large genus.

A related discussion of the quantization of three dimensional gravity (focusing on Einstein gravity rather than Chiral gravity) appears in \cite{noscoop}; we thank these authors for sharing a preliminary version of their paper.  
%We thank the authors of this paper for sharing a preliminary copy of their 

\section{Einstein \& Chiral Gravity}

We will study gravity in three space-time dimensions, with action
\be\label{action}
S = \frac{1}{16\pi G}\int d^3x\sqrt{-g}\left(R+2/\ell^2+{1 \over 2 \mu}
\varepsilon^{\lambda\mu\nu}\Gamma^{r}_{\lambda
\sigma}\left(\partial_{\mu}\Gamma^\sigma_{{r}
\nu}+\frac{2}{3}\Gamma^\sigma_{\mu\tau}\Gamma^\tau_{\nu {r}} \right)\right)~.
\ee
%, and represents the most general diffeomorphism invariant action as all higher derivative terms can be removed by a field redefinition \cite{Gupta:2007th}.
This describes Einstein gravity with a negative cosmological constant and a gravitational Chern-Simons term.
We define the dimensionless couplings
\be\label{couplings}
k_L = { \ell \over 16 G}\left(1+{1\over \mu}\right),~~~~~k_R = {\ell \over 16 G} \left(1-{1\over \mu}\right)~.
\ee
We will set the AdS radius $\ell=1$, so masses and lengths are measured in AdS units.

The simplest case $k_L=k_R$ describes pure Einstein gravity, whose maximally symmetric solution is three dimensional Anti-de Sitter space (AdS$_3$).  There are no local degrees of freedom and every solution is locally AdS.    Nevertheless, the theory contains black holes with horizons and entropies just like their higher dimensional cousins  \cite{Banados:1992wn}.  Our goal is to describe the corresponding microstates.  At first sight this appears impossible, as there are no local degrees of freedom out of which to build black holes.  However, the theory nevertheless possesses global degrees of freedom associated with the topology of space-time which can be quantized.  In particular, we will construct and enumerate black hole microstates with non-trivial topology behind the horizon.  

In order to do so, we will find it convenient to consider the theory with a gravitational Chern-Simons term, in which case $k_L\ne k_R$.  The resulting theory, known as topologically massive gravity \cite{Deser:1981wh, Deser:1982vy}, has a third order equation of motion
\be\label{eom}
{1\over \mu} \epsilon_{\alpha\beta(\mu} \nabla^\alpha {\cal G}_{\nu)}^{\beta}= {\cal G}_{\mu\nu},~~~~~~~{\cal G}_{\mu\nu} \equiv G_{\mu\nu} - {1\over \ell^2} g_{\mu\nu} 
\ee
and possesses solutions which are not locally AdS.  However, when the metric is locally AdS the left hand side of \rref{eom} vanishes and the equations of motion reduce to those of Einstein gravity.  Thus every solution of Einstein gravity can be promoted to a solution of topologically massive gravity.

We wish to study this theory in asymptotically AdS$_3$ space-times, so we must augment the action with a choice of boundary conditions to enforce this condition.  The standard set of boundary conditions are those defined by Brown and Henneaux \cite{Brown:1986nw}.  In this case the diffeomorphisms which act non-trivially on the boundary of AdS$_3$ are not regarded as gauge symmetries but rather as spectrum-generating symmetries which act non-trivially on the states of the theory.  The charges which generate these symmetries obey two copies of the Virasoro algebra, with left and right-moving central charges $c_L=24 k_L$ and $c_R=24 k_R$ \cite{Brown:1986nw, Kraus:2005zm, Henningson:1998gx, Balasubramanian:1999re, deHaro:2000xn, Kraus:2006wn}.  These symmetries act as local conformal transformations on the boundary of AdS$_3$.  This is why every theory of gravity in AdS$_3$ is expected to be dual to a two dimensional conformal field theory (CFT$_2$) with central charges $c_L$ and $c_R$.

In this paper we will focus on the case of chiral gravity, which has $\mu=1$  and $k_R=0$ \cite{Li:2008dq}.  In this case the right-moving Virasoro charges vanish: the right-moving conformal symmetries become gauge transformations and the theory possesses only chiral (left moving) degrees of freedom.  The quantization of the gravitational Chern-Simons term implies that $k_L=k\in \ZZ$, which can be viewed as a consequence of the fact that the boundary dual is a modular invariant chiral CFT with central charge $c_L=24 k$ \cite{Witten:2007kt}. 

Remarkably, once back-reaction is taken into account all known propagating wave solutions turn out to violate the boundary conditions; there is a large literature on this subject, so we refer the reader to \cite{Maloney:2009ck} and references therein for an extensive discussion.  All known, finite energy solutions of chiral gravity are locally AdS$_3$.\footnote{Solutions which are not locally AdS are known  \cite{Compere:2010xu}, but these appear to have infinite energy so will not concern us here.}  We will therefore proceed to consider just the locally AdS solutions of chiral gravity.

% \section*{}
\section{Classical Solutions}
\subsection{AdS$_3$ and BTZ}

The classical solutions of gravity in AdS$_3$ are labelled by an energy $\Delta$ and a spin $J$. In the language of the dual CFT, $\Delta=L_0+{\bar L}_0$ is the scaling dimension of the corresponding CFT state and $J = L_0 - {\bar L}_0$.  
In Einstein gravity a static solution has $J=0$.  
In chiral gravity, however, the mass and spin of a solution are altered due to the presence of the Chern-Simons term  \cite{Deser1982}.  In particular, since ${\bar L}_0=0$ all states have mass equal to spin $\Delta=J$.  In chiral gravity even a static solution will have spin $J\ne 0$ (see e.g. \cite{Deser:1989ri}).

The simplest solution is AdS$_3$, which has topology $D_2\times \RR$ where $D_2$ is a disk.  This geometry is the minimum energy ground state of the theory.  It will be convenient to shift our energies so that empty AdS has mass $\Delta=-k$ rather than $\Delta=0$.  In the CFT language, $\Delta=-k=-{c/ 24}$ is the energy of the vacuum state on the cylinder.
The metric on AdS can be written in FRW coordinates as
\be\label{metric}
ds^2 = -dt^2 + \cos^2 t ~d\Sigma^2 
\ee
where $d\Sigma^2$ is the constant negative curvature (Poincar\'e) metric on the disk $D_2$.  This somewhat unfamiliar coordinate system covers only a portion of the global AdS$_3$ space-time, as shown in Figure 1.  The conformal boundary of AdS$_3$ is the cylinder $S_1\times \RR$; the boundary of the disk $\partial D_2=S^1$ intersects this conformal boundary on a circle.

\begin{figure}\label{f1}
\centering
\includegraphics[scale=0.6, clip=true]{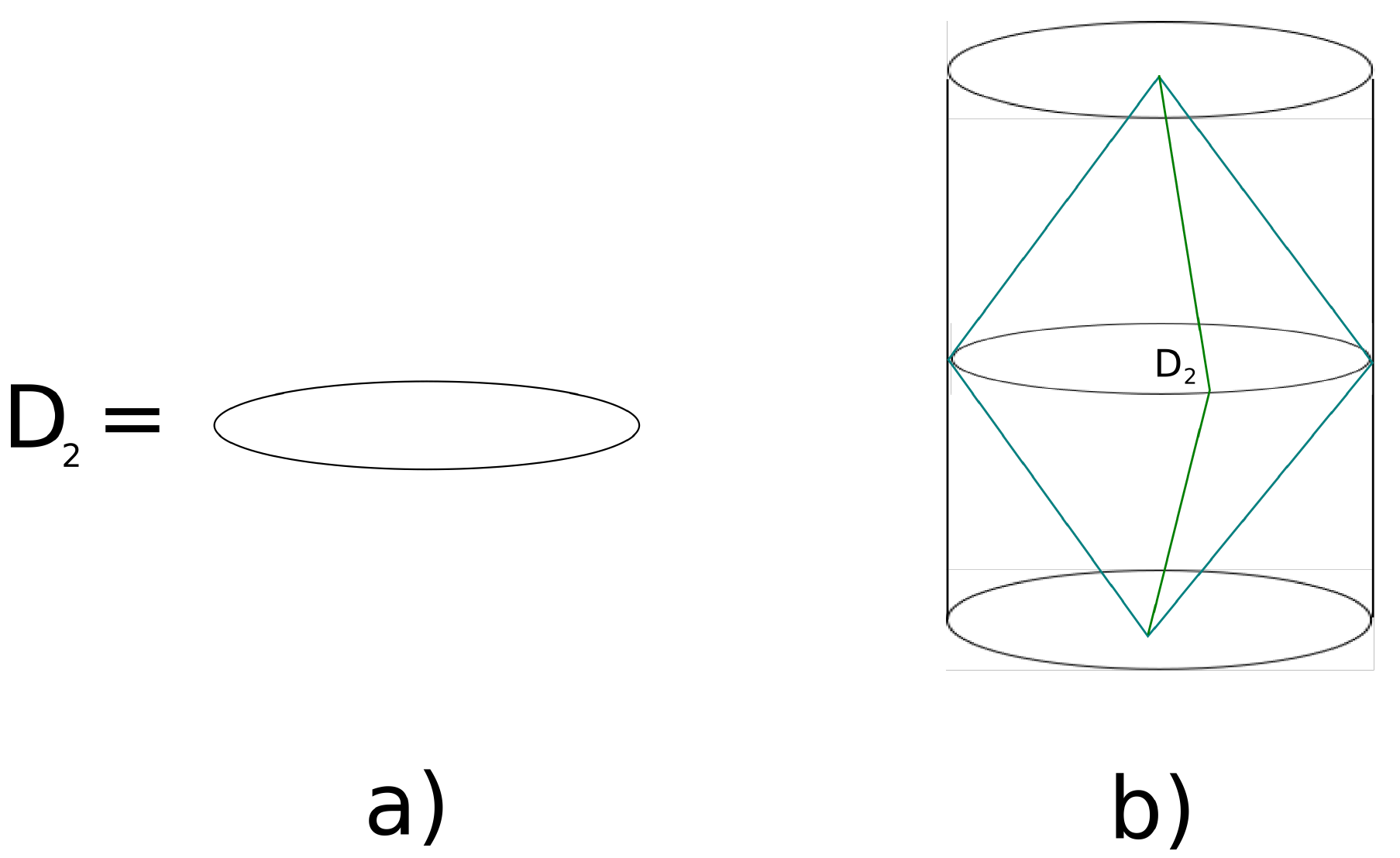}
%~~~~~~~
%\includegraphics[scale=0.9,clip=true]{}  
         \caption{AdS is the solid cylinder whose constant time slices are given by the disk $\Sigma=D_2$ in (1a).  The metric \rref{metric} covers the coordinate patch of AdS inside the green diamond depicted in (1b).}
\end{figure}

More general solutions will have topology $\Sigma\times \RR$ for some surface $\Sigma$, where $\RR$ can be viewed as a time coordinate and the surface $\Sigma$ as a constant time slice.
These solutions can be constructed by quotienting AdS$_3$ by a discrete subgroup $G$ of its $SO(2,2)$ isometry group.  
 A simple class of such solutions is given again by \rref{metric}, where $d\Sigma^2$ now represents the constant negative curvature metric on a Riemann surface $\Sigma$.  A detailed description of these solutions is contained in e.g. \cite{Aminneborg:1997pz, Brill:1998pr, Brill:1999xm, Krasnov:2000zq}; we summarize here only a few important features.

The simplest non-trivial case is when the spatial slice $\Sigma$ has the topology of the annulus.  This geometry is a quotient of AdS$_3$ by a single hyperbolic element of $SO(2,2)$.  In this case the metric is precisely that of the static BTZ black hole, albeit in a somewhat unfamiliar coordinate system.  This is shown in Figure 2.  

\begin{figure}\label{f1}
\centering
\includegraphics[scale=0.8, clip=true]{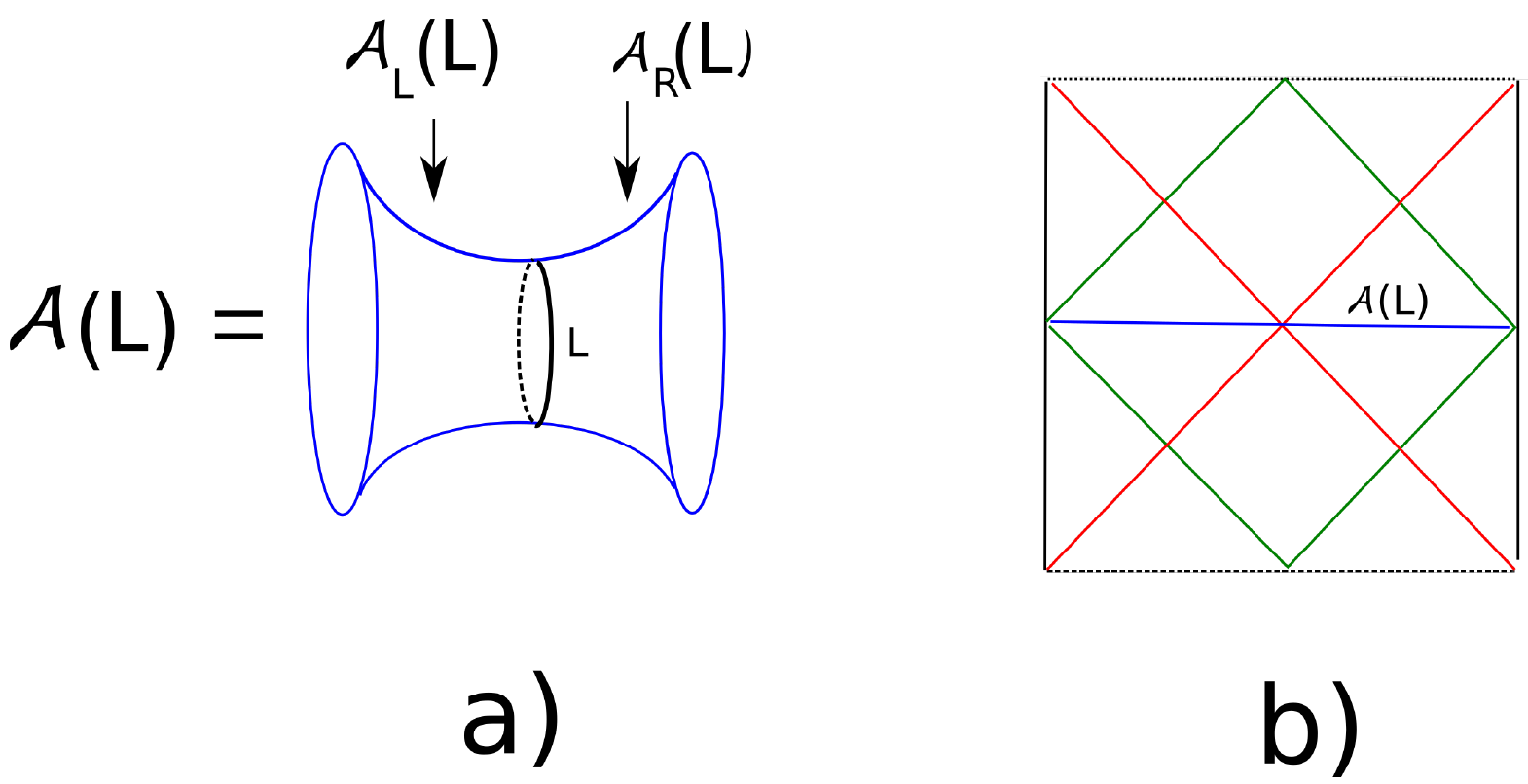}
%        ~~~~~~~
%         \includegraphics[scale=0.9,clip=true]{}
        \caption{The constant time slices of the BTZ black hole are the annulus, whose constant time slice is shown in (2a).  The Penrose diagram of BTZ is shown in (2b). The $t=0$ slice, shown in blue, is the annulus.  The metric \rref{metric} covers the coordinate patch inside the green diamond.  The event horizon is shown in red, and intersects the annulus at the minimum length geodesic of length $L$. Dotted lines indicate past and future singularities.} 
\end{figure}

The constant negative curvature metric on the annulus has a single free parameter, which we denote $L$, which is the length of the minimum length geodesic around the annulus.   We will denote the annulus with a geodesic of  length $L$ as $\A(L)$.  We can cut this annulus into two components along this geodesic: 
\be
\A(L) = \A_R(L) \cup \A_L(L)~,
\ee where $\A_{R,L}(L)$ describe the two halves of the annulus on the right and left sides of the geodesic, as in Figure (2a). %either side of the geodesic.

The two boundaries of the annulus intersect the two asymptotic conformal boundaries of the BTZ geometry.  At $t=0$, the surfaces $\A_{R}(L)$ and $\A_L(L)$ describe the zero time slices of the right and left asymptotic regions of the BTZ black hole, respectively.  The geodesic is the Einstein-Rosen bridge connecting these two regions, and $L$ is precisely the area (i.e. length) of the black hole event horizon.   The mass and entropy of the BTZ black hole can be computed in chiral gravity, taking into account the gravitational Chern-Simons term \cite{Kraus:2005zm, Moussa:2003fc, Solodukhin:2005ah, Sahoo:2006vz, Park:2006gt, Tachikawa:2006sz}.
The black hole has mass
\be
\Delta =  {1\over 4 \pi^2} k L^2
\ee
and Bekenstein-Hawking entropy
\be
S_{BH} = 4\pi \sqrt{k \Delta}~.
\ee
%This is not just the Bekenstein-Hawking entropy \cite{}, but includes the corrections due to Wald \cite{} that arise in the presence of a gravitational Chern-Simons term.

\subsection{Microstate Geometries}

The BTZ black hole describes a state with two asymptotic boundaries.  We wish to describe configurations with only one asymptotic boundary.  One family of such geometries -- which we refer to as microstate geometries -- is again given by \rref{metric}, where $d\Sigma^2$ now represents the hyperbolic metric on a smooth surface $\Sigma_g$ of genus $g\ge1$.  We will demand that $\Sigma_g$ has one hole, so that  $\partial \Sigma_g=S^1$.  This ensures that the geometry has only one asymptotic boundary, which intersects the boundary of the Riemann surface $\Sigma_g$ along this circle. We will refer to these as ``microstate geometries."  A sample microstate geometry with $g=1$ is shown in Figure 3.

\begin{figure}\label{f1}
\centering
\includegraphics[scale=0.65, clip=true]{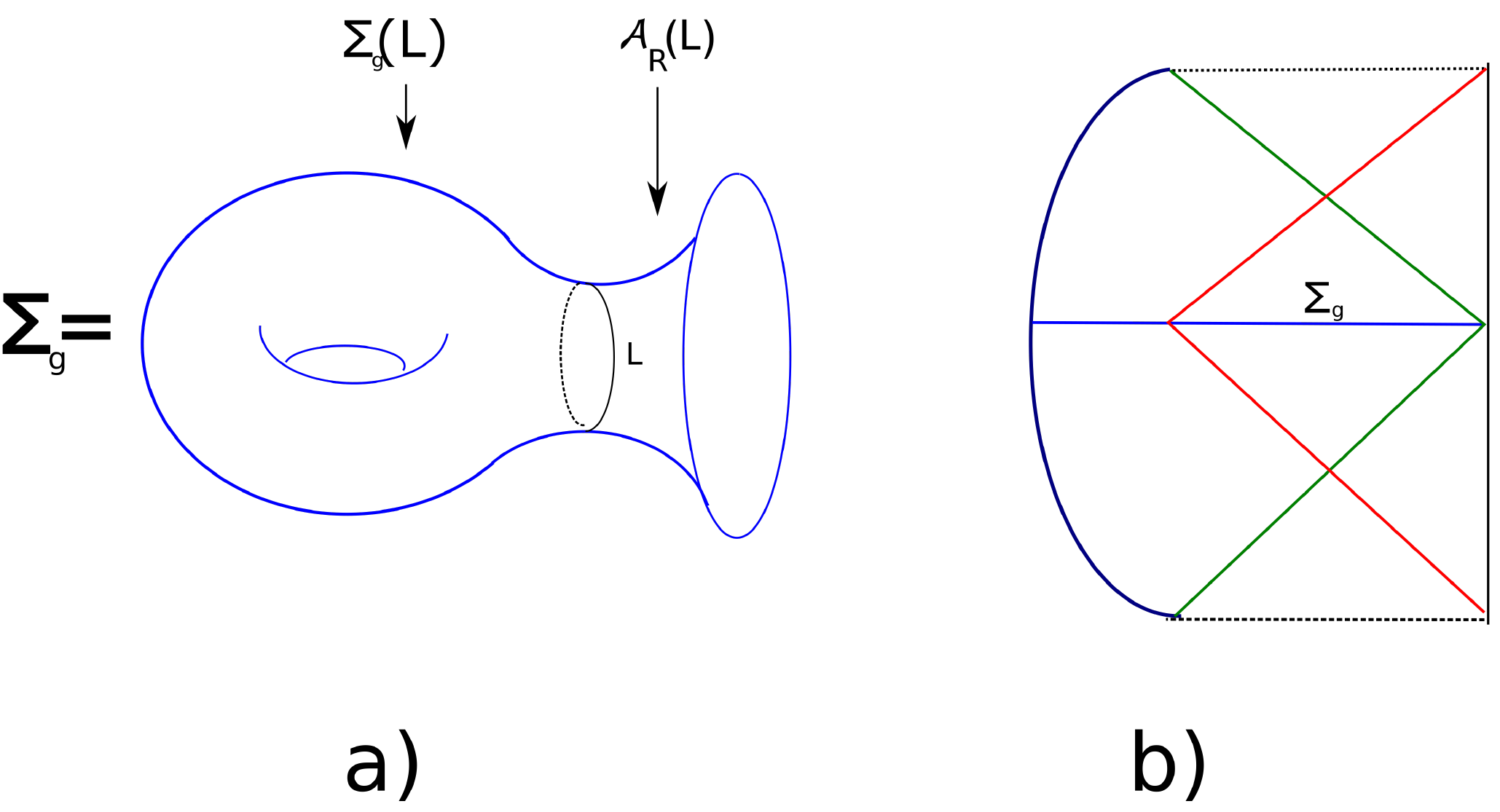}
%        ~~~~~~~
%         \includegraphics[scale=0.9,clip=true]{}
 \caption{A microstate geometry whose constant time slice is the surface $\Sigma_g$ with one hole (3a). The Penrose diagram is sketched in (3b).  The blue line is the surface $\Sigma_g$ at the $t=0$ slice.  The metric \rref{metric} covers the coordinate patch inside the green diamond. The event horizon is shown in red, which intersects $\Sigma_g$ at the geodesic of length $L$. The curved line on the left side of 3b) indicates that the interior geometry caps off smoothly.  This is only a sketch of the Penrose diagram; since the geometry breaks the $U(1)$ rotation symmetry of AdS$_3$, one cannot draw a two dimensional Penrose diagram inside the horizon.}
\end{figure}

At fixed genus $g$ there are many different choices of surface $\Sigma_g$, which are labelled by a set of continuous parameters (moduli).  
The most important modulus is the length $L$ of the minimum length geodesic which separates the asymptotic boundary from the the topologically non-trivial part of $\Sigma_g$.   It is useful to think of the surface $\Sigma_g$ as cut into two pieces along this geodesic:
\be
\Sigma_g = {\Sigma}_g(L)\cup \A_R(L)~
\ee 
as shown in Figure 3a.
Here ${\Sigma}_g(L)$ is a bordered Riemann surface with a single geodesic boundary of length $L$, and $\A_R(L)$ is the right half of the hyperbolic annulus.  

The Cauchy development (both forwards and backwards in time) of $\A_R(L)$ describes one of the asymptotic regions outside the horizon of the BTZ black hole. 
This means that a microstate geometry has an event horizon of length $L$, and that the geometry outside this event horizon is {\it identical} to that of a BTZ black hole with horizon area $L$.  In other words, the metric in the right quadrant of Figure  3b is the same as that in the right quadrant of Figure 2b.  They both describe the region outside the horizon of a BTZ black hole.  Thus to an asymptotic observer near the boundary of AdS$_3$ a microstate geometry is classically indistinguishable from a black hole.  

These microstate geometries can be described more explicitly as quotients of AdS$_3$ by a discrete subgroup $G$ of its $SO(2,2)$ isometry group.  If we write $SO(2,2)=SL(2,\RR)\times SL(2,\RR)$, then the diagonal $SL(2,\RR)_{diag} \subset SL(2,\RR)\times SL(2,\RR)$ are the isometries of the Poincar\'e disk $D_2$ which is the constant $t$ slice of AdS$_3$.  The microstate geometries are quotients of AdS$_3$ by a freely acting group $G\subset SL(2,\RR)_{diag}$ sitting inside this diagonal subgroup.
Since $D_2$ is topologically trivial, the quotient group $G$ is precisely the fundamental group $\pi_1(\Sigma_g)$ of the spatial surface. We will restrict our attention to smooth solutions, so that the elements of $G$ are hyperbolic elements of $SL(2,\RR)$; elliptic and parabolic elements would lead to surfaces with conical deficit and cusp singularities, respectively.

The fact that $G$ lives in the diagonal subgroup $SL(2,\RR)_{diag}$ reflects the fact that all of these geometries have a time reflection $t\to -t$ symmetry; the two copies of $SL(2,\RR)$ are exchanged by this parity operation.  More general solutions, describing rotating solutions without this symmetry, would be quotients by more complicated subgroups $G\subset SO(2,2)$.  We will argue below that in chiral gravity it is sufficient to consider just the diagonal quotients.  The non-diagonal quotients, which would be relevant in Einstein gravity, are much more complicated.

\subsection{Microstates vs. Macrostates}

Although classical indistinguishable, the quantum interpretation of the microstate geometries is very different from that of the BTZ geometry.  
 The BTZ geometry should be regarded as a mixed state which is obtained by coarse-graining over a family of black hole microstates \cite{Strominger:1997eq, Martinec:1998wm, Maldacena:1998bw, Maldacena:2001kr}. 
 The microstate geometries  are regarded as pure quantum states for gravity in AdS$_3$, rather than 
 mixed states 
%In fact, Maldacena \mcite{Maldacena:2001kr}, who in the same paper he noted that 
 \cite{Krasnov:2000zq, Maldacena:2001kr}.  Other investigations of the microscopic structure of these geometries include \cite{Horowitz:1998xk, Krasnov:2003ye, Maldacena:2004rf, Freivogel:2005qh, Skenderis:2009ju, Balasubramanian:2014hda, Guica:2014dfa, Marolf:2015vma, Bao:2015bfa}.  

The pure state interpretation of these geometries is easiest to see when we consider the CFT duals of these states.  Let us begin with the BTZ black hole, which has two asymptotic boundaries.  The BTZ black hole is described by a state in the Hilbert space which is a tensor product of two copies of the CFT Hilbert space.  This state can be constructed by analytically continuing to Euclidean signature, where -- following Hartle and Hawking \cite{Hartle:1983ai} -- the wave function on the $t=0$ slice is given by a Euclidean path integral.  This leads to a particular pure state which contains entanglement between the two asymptotic boundaries.  This pure state in the doubled CFT Hilbert space is known as the thermofield double state.  It has the property that tracing out over one of the copies of the Hilbert space leads to a thermal density matrix in the other copy.  Thus an observer who only has access to one CFT -- such as an observer who lives in the right-hand quadrant of the BTZ geometry in Figure 2b -- will see a thermal mixed state with temperature equal to the Hawking temperature.  

The microstate geometries, on the other hand, have only one asymptotic boundary.  The Euclidean path integral leads to a specific construction of a pure state in the single CFT Hilbert space.  The state is defined by a path integral on the surface $\Sigma_g$ with one hole.  In particular, we can regard the CFT path integral on $\Sigma_g$ as a functional of CFT data on the boundary $\partial \Sigma_g = S^1$.  This defines a state in the Hilbert space of the CFT on the cylinder.  The moduli of $\Sigma_g$ continuously parameterize a basis of states in the CFT.  
%This generalizes the BTZ case, where the state is given by a path integral on the torus which is the boundary of the Euclidean black hole geometry.  
The details of this construction are rather complicated but not particularly relevant for this paper; we refer the reader to the literature for a detailed discussion of these states.  

We will proceed by assuming that microstate geometries construct pure states rather than mixed states.
This is essentially the assumption that we have a ``pure" theory of quantum gravity, i.e. a theory with only geometric degrees of freedom described by spaces of smooth metrics.  Of course, it may turn out that in order to correctly reproduce black hole entropy one must include additional degrees of freedom.
We note that the microstate geometries are regarded as pure states even though they have an event horizon; this differs, for example, from the fuzzballs of \cite{Mathur:2005zp}. 

\section{The Phase Space of Solutions}

\subsection{Configuration Space of Black Hole Microstates}

The microstate geometry with spatial slice $\Sigma_g$ has many continuous moduli, which parameterize the moduli space of constant negative curvature metrics on $\Sigma_g$.\footnote{There is a vast literature on moduli of Riemann surfaces.  Many useful facts can be found in \cite{harris1998moduli} and in two more recent reviews \cite{wolpert, MR3184165}.}  To proceed, we will use the fact that $\Sigma_g$ is the union of 
${\Sigma}_g(L)$, the bordered Riemann surface with a single geodesic boundary of length $L$, 
and $\A_R(L)$, the half-annulus with one geodesic boundary of length $L$ and one asymptotic boundary.
The total moduli space of $\Sigma_g(L)$ can be decomposed into two parts
\be
\M^{total} = \M_{g,1}(L) \oplus \M^{bdy}(L)
\ee
where $ \M_{g,1}(L) $  is the moduli space of bordered Riemann surfaces and $\M^{bdy}(L)$ is the moduli space of $\A_R(L)$.

Let us first consider $ \M^{bdy}(L)$.  The surface $\A_R(L)$ is an annulus, with one boundary a geodesic of length $L$ and the other infinite length boundary matching onto the asymptotic boundary of AdS$_3$.  The geometry of this surface is uniquely determined by the length $L$ of the geodesic, so it might appear that $ \M^{bdy}(L)$ is  trivial.  However, this is not quite the case.  When we impose Brown-Henneaux boundary conditions for AdS$_3$ gravity, the diffeomorphisms that act non-trivially on the boundary of AdS are not regarded as pure gauge.  So when we consider the space of metrics on $\A_R(L)$ we should regard two metrics as identical only if they are related by a diffeomorphism which vanishes sufficiently quickly near the asymptotic boundary of $\A_R(L)$.  The moduli space $ \M^{bdy}(L)$ is therefore $diff(S^1)$, the group of non-trivial diffeomorphisms acting at asymptotic infinity.  Two metrics on $\A_R(L)$ which describe the same geometry, but differ by an element of $diff(S^1)$, describe two different points in $ \M^{bdy}(L)$.  Closely related considerations appeared in \cite{Maloney:2007ud}, who considered the case where the surface was a disk rather than a half-annulus.

The appearance of $diff(S^1)$ as (part of) the configuration space of classical gravity has a very natural CFT interpretation.  The generators of $diff(S^1)$ are the Virasoro generators of local conformal transformations in the boundary CFT.   
The quantization of $diff(S^1)$ leads to representations of the Virasoro algebra (see e.g. \cite{Witten:1987ty}).  The states that one obtains from this quantization are the perturbative graviton states of AdS$_3$, known as boundary gravitons (see e.g. \cite{Maldacena:1998bw, Witten:2007kt, Maloney:2007ud}. These states do not describe local degrees of freedom, but rather excitations that exist only in the presence of a boundary, much like  edge states in a quantum hall system.  

This leads to the following appealing picture: a state which one obtains from the quantization of ${\Sigma}_g(L)$, the behind-the-horizon geometry, is interpreted as a primary state in the CFT.  The descendant states, built by acting with Virasoro raising operators on this primary, arise from the quantization of $\A_R(L)$.  In other words, the quantization of $\Sigma_g(L)$ leads to a black hole microstate, and the quantization of $\A_R(L)$ dresses this black hole with boundary gravitons.
 
We now turn to the moduli space $\M_{g,1}(L)$ of the geometry behind the horizon.  A simple geometric description of $\M_{g,1}(L)$ is as follows.  We first decompose $\Sigma_g(L)$ into a collection of pairs of pants sewn together along their cuffs.   Each pair of pants is a sphere with three holes (cuffs), each of which is a geodesic.  The constant negative curvature metric on a pair of pants is uniquely specified by the lengths of the geodesics around each cuff.  To determine the metric on $\Sigma_g(L)$, we  must therefore specify a length $L_i$ for each cuff, along with a twist parameter $\tau_i$ describing the relative angle at which these cuffs are sewn together.   The number of internal cuffs is $3g-2$, so a point in $\M_{g,1}(L)$ is described by $6g-4$ parameters.  So the moduli space $\M_{g,1}(L)$ has complex dimension $3g-2$.  The $(L_i,\tau_i)$  are known as Fenchel-Nielsen coordinates on moduli space.

To completely determine the metric on $\Sigma_g(L)$ we must also specify the length of the geodesic boundary $L$.  There is  a twist parameter $\tau$ associated with the length $L$, which describes how the black hole interior is matched onto asymptotic infinity; $\tau$ is just the usual angular coordinate in the asymptotic region of AdS$_3$.

The parameters $(L_i,\tau_i)$ described above parameterize the Teichm{\"u}ller space $T_{g,1}(L)$ of bordered Riemann surfaces.  It is important to note that many points in $T_{g,1}(L)$ describe the same Riemann surface.  For example, the twist parameters $\tau_i$ are periodic with $\tau_i\sim\tau_i+2\pi$, so the moduli space $\M_{g,1}(L)$ is actually a quotient of Teichm{\"u}ller space.  The surface $\Sigma_g(L)$ will have many different pair-of-pants decompositions, and we must perform the corresponding twist parameter quotient for each possible pair of pants decomposition.  The result is that the moduli space is a complicated quotient of Teichm{\"u}ller space: $\M_{g,1} (L)= T_{g,1}(L)/\Gamma$, where $\Gamma$ is the mapping class group of the surface $\Sigma_g(L)$.

The mapping class group action has fixed points, which make $\M_{g,1}(L)$ an orbifold rather than a manifold.  The fixed points of $\Gamma$ are surfaces with automorphisms, and the stabilizer of a point is the corresponding automorphism group.  
%These points should be treated carefully in a detailed quantization of $\M_{g,1}(L)$, which we hope to return to in the future, but for our present purposes they will not be important.

%We will comment on this below when we discuss the quantization of these geometries.
%These points must be treated carefully in a detailed quantization of $\M_{g,1}(L)$, which we hope to return to in the future.}

%\begin{figure}
%\centering
%%        \includegraphics[scale=0.8,viewport=0 440 400 570,clip=true]{}
%\caption{The pair of pants decomposition for a surface $\Sigma_g$, which is obtained by gluing together pairs of pants along the cuffs depicted in blue.}
%\end{figure}
% 

\subsection{The Symplectic Structure \& Chern-Simons Theory}

We now wish to quantize the space of classical solutions and describe the corresponding quantum states.  We first note that each point in the classical configuration space of solutions can be regarded as a point in phase space.  So the moduli space $\M_{g,1}(L)$ described above should be regarded as a phase space.  We need to determine the symplectic structure on this phase space.

%The simplest way to determine this symplectic structure is to use 
To do so, we will use the Chern-Simons formulation of three dimensional gravity  \cite{Achucarro:1987vz, Witten:1988hc}, where the action \rref{action} is written as 
\be\label{cs}
S = k_L I_{CS}[A_L] - k_R I_{CS}[A_R]
\ee
where $A_{R,L}$ are a pair of $SL(2,\RR)$ Chern-Simons gauge connections and $I_{CS}$ the corresponding Chern-Simons actions.  Although the equations of motion of topologically massive gravity differ from those of Chern-Simons theory (the latter possesses no local degrees of freedom) they are identical for the locally AdS solutions.  So the Chern-Simons formulation provides a simple way of determining the appropriate symplectic structure for the solutions of interest.

It is important to note that we are not asserting here that $SL(2,\RR)\times SL(2,\RR)$ Chern-Simons theory is identical to three dimensional gravity as a quantum theory.  For example, the definition of Chern-Simons theory requires us to fix a topological three-manifold as a background, whereas in gravity it is natural to consider space-times of different topology.  Instead, our approach is to first divide up the phase space of gravity into sectors of fixed topology, and then to use Chern-Simons theory as an aid in the quantization of a particular topological sector.  A similar strategy was used in \cite{Castro:2011xb} in the study of de Sitter quantum gravity.\footnote{In a related point, we note that we have been sloppy about which global form of the $SO(2,2) = SO(2,1)\times SO(2,1)$ gauge group we are considering.  Instead of $SL(2,\RR)$, the two-fold cover of $SO(2,1)$, one could consider a Chern-Simons theory based on a different gauge group.  This could lead to phase spaces which are different quotients of Teichm{\"u}ller space.  We will rely here on the geometric picture, where it is clear that the phase space of microstate geometries is $\M_{g,1}$, though other choices might lead to interesting results.  Related points were discussed in \cite{Witten:2007kt}.}
%Moreover, we should emphasize that in principle the 

The microstate geometries can be described in Chern-Simons language a follows.  
%The Chern-Simons equation of motion say that $A_L$ is a flat, $SL(2,\RR)$ connections.
A microstate geometry has topology $\Sigma_g\times \RR$.  A solution of Chern-Simons theory on a manifold of this topology is given by a pair of flat $SL(2,\RR)$ connections on the surface $\Sigma_g$. 
A flat $SL(2,\RR)$ connection is uniquely specified (up to gauge transformation) by its holonomies around the topologically non-trivial cycles of $\Sigma_g$.   These holonomies are characterized by a pair of maps 
\be
hol_{L,R}: \pi_1(\Sigma_g)\to SL(2,\RR)~
\ee 
associating an element $(hol_{L}(\gamma),hol_R(\gamma)) \in SL(2,\RR)\times SL(2,\RR)$ to each topologically nontrivial cycle $\gamma$ of $\Sigma_g$.  This holonomy data is precisely what we need to construct a microstate geometry as a quotient of AdS$_3$ by a discrete group $G$.  In particular, the fundamental group $\pi_1(\Sigma_g)$ is just the quotient group $G$, and the holonomy maps $hol_{L,R}$ provide an embedding of this group into the $SL(2,\RR)\times SL(2,\RR)$ isometry group of AdS$_3$.  Since we are restricting our attention to smooth geometries, the holonomies of this connection are hyperbolic.

This family of Chern-Simons solutions is more general than the microstate geometries constructed earlier.  The microstate geometries described above have $G\subset SL(2,\RR)_{diag}\subset SL(2,\RR)\times SL(2,\RR)$, so $hol_L(\gamma)=hol_R(\gamma)$.
The solutions with $hol_L(\gamma)\ne hol_R(\gamma)$  describe solutions which do not have a time reversal $t\to -t$ symmetry, such as rotating black holes.  

In the case of Chiral gravity, however, $k_R=0$ and we are left with a single dynamical $SL(2,\RR)$ Chern-Simons field $A_L$.   The other gauge field $A_R$ can be chosen to be whatever we like.  We will simply choose to set $A_L=A_R$.  This means we are free to consider only the time-reversal symmetric microstate geometries described above.  In other words, in chiral gravity the set of solutions is described by the moduli space $M_{g,1}(L)$.

We can now use previous results to compute the symplectic structure on the moduli space of flat connections.  In the case of $SL(2,\RR)$ Chern-Simons theory, the symplectic structure has been computed \cite{Verlinde:1989ua, Killingback:1990hi}.  It is
\be
\omega = {k\over \pi}~ \omega_{WP} (\M_{g,1}(L))
\ee
where $\omega_{WP} (\M_{g,1}(L))$ is the Weil-Petersson symplectic structure on $\M_{g,1}(L)$.  
In the coordinates $(L_i, \tau_i)$ on $\M_{g,1}(L)$ described above, the Weil-Petersson symplectic structure can be written as  \cite{wolpert1981elementary}\footnote{We have normalized our twist parameters to be periodic with period $2\pi$, since $\tau$ is the usual angular coordinate at infinity. In the literature $\tau_i$ is usually taken to be periodic with period $L_i$, leading to a slightly different formula for $\omega_{WP}$.}
\be\label{omega}
\omega_{WP}  (\M_{g,1}(L)) = {1\over 2\pi} \left( L\ d L \wedge d\tau + \sum_i L_i dL_i \wedge d \tau_i \right)~.
\ee
Here $i$ runs over the cuffs which lie behind the horizon and $(L,\tau)$ are the length and twist parameters of the horizon.  

%{\bf check these formulas, factors of 2}

\subsection{Einstein Gravity vs. Chiral Gravity}

Before proceeding, it is useful to contrast the case of chiral gravity with that of Einstein gravity.  The microstate geometries we are considering have a time reflection symmetry: $t\to -t$.  Their configuration space $\M_{g,1}(L)$ is the space of metrics on the spatial slice at time $t=0$.  The extrinsic curvature of this slice vanishes.  In Einstein gravity, the metric on the spatial slice $\Sigma$ is canonically conjugate to the extrinsic curvature of $\Sigma$.  This means that in Einstein gravity the symplectic structure on the configuration space of microstate geometries would be zero. The gravitational Chern-Simons term, however, turns this configuration space into a genuine phase space with non-zero symplectic structure. 

The situation is somewhat analogous to that of particles moving in magnetic field, where the configuration space of positions on the $(x,y)$ plane becomes a phase space with $[x,y]= B$ in the presence of  a magnetic field $ {\bf B} = B {\hat z}$.  The quantum states which one obtains by quantizing this phase space are Landau levels.  For us, the gravitational Chern-Simons term plays the role of the magnetic field, in that it promotes the configuration space $\M_{g,1}$ to a phase space.  Our goal is  to count the corresponding Landau levels.

Before doing so, we will make one comment on the global structure of the phase space of microstate geometries.  In chiral gravity, one studies the space of holonomies living in $SL(2,\RR)$, which are parameterized by points in the Teichm{\"u}ller space  $T_{g,1}$.  When we interpret these solutions as microstate geometries, it is clear that we should in fact quotient this Teichm{\"u}ller space by the mapping class group to obtain the moduli space $\M_{g,1}(L)$.  Global diffeomorphisms are used to turn the a-priori infinite Teichm{\"u}ller space into a finite moduli space, which one has a hope of quantizing to obtain a finite number of states.  In  Einstein gravity one must instead study the full space of holonomies living in $SL(2,\RR)\times SL(2,\RR)$.  Locally this looks like two copies of Teichm{\"u}ller space.  However, even after accounting for global diffeomorphisms this space is quotiented out by only one copy of the mapping class group, rather than two.
So in Einstein gravity it appears difficult to ever obtain a finite Hilbert space by quantizing the phase space of geometries.  Chiral gravity seems like the only scenario where one can obtain a discrete spectrum with a finite number of microstates by quantizing geometry.

\section{Counting Black Hole Microstates}

\subsection{Quantizing Phase Space}

Now that we have a symplectic structure we can attempt to quantize our phase space.
The quantization of the moduli space of Riemann surfaces is a rich mathematical subject, a complete discussion of which is beyond the scope of this paper (see e.g. \cite{fock1997dual, Kashaev:1998fc, Chekhov:1999tn, Teschner:2003em, Teschner:2003at, Teschner:2005bz, Teschner:2014vca}).   We will make only general remarks with the goal of estimating the number of black hole microstates, leaving a detailed characterization of their structure to future work.

When the microstate geometries are quantized, the coordinates $(L_i,\tau_i)$ on moduli space $\M_{g,1}(L)$ are promoted to operators acting on a Hilbert space.  In particular, the area $L$ of the event horizon becomes an operator, whose spectrum we wish to compute.  We first note that, from \rref{omega}, the area operator $L$  is canonically conjugate to the twist parameter $\tau$, so
\be
[L^2,\tau] =  i {4\pi^2 \over k}~.
\ee
Since $\tau$ is periodic with period $2\pi$, the area operator has quantized spectrum:
\be\label{delta}
\Delta =  {1\over 4 \pi^2} k L^2 \in \Z~.
\ee
%The black hole horizon area is quantized.
Although the quantization of the horizon area might appear impressive, it has a simple explanation: in a chiral theory every state has mass $\Delta$ equal to its spin $J$, so \rref{delta} is  the usual quantization of angular momentum.   
%Note that we have found half-integer quantization because we are working with $SL(2,\RR)$ 

Before discussing the quantization of this phase space, we first note that the moduli space $\M_{g,1}(L)$ is non-compact -- it has boundaries, where cycles of the surface $\Sigma_g(L)$ pinch.  The most convenient way to deal with these boundaries is to use the Deligne-Mumford compactification of moduli space, which amounts to a particular prescription for how boundary components are included in $\M_{g,1}(L)$ as cycles pinch.  The advantage of this approach -- which we will use -- is that it turns the moduli space into a projective variety, to which one can apply the techniques of algebraic geometry.\footnote{In the present work, where we are content to estimate the volumes of the relevant phases spaces, subtleties about the choice of compactification are likely unimportant.}
%In the present paper we will assume that this is the correct approach, though this subject should be revisited in the future. 

The space $\M_{g,1}(L)$ is  quite complicated.  
Fortunately,  Mirzakhani recently made significant progress in the computation of the volume of $\M_{g,1}(L)$ with respect to $\omega_{WP}$ \cite{Mirzakhani:2007aa}.  This will allow us to estimate the corresponding number of quantum states.  Although we could work directly with the moduli space of $\M_{g,1}(L)$ of bordered Riemann surfaces, we will find it convenient to rephrase these results in the language of punctured surfaces, following \cite{MR2257394}.   

We begin with the fact that the moduli space $\M_{g,1}(L)$ of bordered Riemann surfaces (i.e. surfaces with a single geodesic boundary) is symplectomorphic to the moduli space of punctured Riemann surfaces, denoted $\M_{g,1}$ \cite{MR2257394}.  Under this map, the geodesic boundary of $\Sigma_g(L)$ with length $L$ has been shrunk to a point.  Crucially, under this symplectomorphism the symplectic structure of $\M_{g,1}(L)$ does not map to the Weil-Petersson symplectic structure of $\M_{g,1}$.  Rather, they are related by
%\footnote{A nice review of this work appears in \cite{MR3184165}.}
\be\label{symplecto}
\omega_{WP}(\M_{g,1}(L)) =  \omega_{WP}  + {1\over 2} L^2 \psi_1
\ee
where $\omega_{WP}$ is the Weil-Petersson symplectic structure on  $\M_{g,1}$ and $\psi_1$ is a correction term defined as follows.
On a once-punctured Riemann surface the cotangent bundle at the puncture provides a natural line bundle over $\M_{g,1}$. $\psi_1$ is the curvature of this bundle. 

We can now approximate the number of genus $g$ microstates with energy $\Delta$ as the volume of $\M_{g,1}$ with respect to this symplectic structure:
\bea\label{ng}
N_g(k,\Delta) &\approx&
{ 1\over (3g-2)!} \int_{\M_{g,1}} \left({\omega\over 2\pi}\right)^{3g-2} \cr
&=& { 1\over (3g-2)!} \int_{\M_{g,1}}  \left({{k \over 2\pi^2}  {\omega_{WP}}+ \Delta \psi_1}\right)^{3g-2} 
\eea 
%Where $ \omega_{WP}$ is the Weil-Petersson form on $\M_{g,1}$.  
We emphasize that this differs from the usual Weil-Petersson volume of $\M_{g,1}$ due to the correction term $\psi_1$; 
it is precisely through this correction term that the area of the black hole horizon enters the computation.

This formula has a very nice interpretation if we regard the symplectic structure as an element of cohomology $[\omega]$ on $\M_{g,1}$. In this case \rref{ng} can be expressed in terms of intersection numbers on $\M_{g,1}$.  In particular, the Weil-Petersson symplectic structure  is proportional to a class 
$\kappa_1$ on $\M_{g,1}$ known the first tautological class: $[\omega_{WP}] = 2 \pi^2 \kappa_1$  \cite{MR727702}.  We can therefore rewrite \rref{ng} as %that the number of genus $g$ microstates can be computed in terms of intersection numbers:
\begin{eqnarray}\label{approx}
N_g(k,\Delta) \approx %{1\over (3g-2)!} \int_{\M_{g,1}}  \left({k \kappa_1 + \Delta \psi_1}\right)^{3g-2} = 
\sum_{d=0}^{3g-2} { k^{3g-2-d} \Delta^d \over (3g-2-d)!~ d!}~\int_{\M_{g,1}} \kappa_1^{3g-2-d}  \psi_1^{d}~.
\end{eqnarray}
%We note that area of the black hole horizon $\Delta$ enters only through the coefficient $\Delta$ of this correction term.
%
%$\omega_{k,\Delta} \equiv k \kappa_1 + \Delta \psi_1 $
%
%where $\kappa_1$ is the Weil-Peterson symplectic structure on $\M_{g,1}$ and $\psi_1$ is a correction term defined as follows.
%On a once-punctured Riemann surface, the cotangent bundle at the puncture provides a natural line bundle over $\M_{g,1}$.  $\psi_1$ is the first Chern class of this bundle.  
%The volume of $\M_{g,1}(L)$ is then related to the intersection numbers between the  $\kappa_1$ and $\psi_1$ classes.  
%
%
%
%
%\begin{eqnarray}\label{approx}
%N_g(k,\Delta) &\approx& \int_{\M_{g,1}}  e^{k \kappa1 + \Delta \psi_1} = \sum_{d=0}^{3g-2} { k^{3g-2-d} \Delta^d  \over (3g-2-d)!~ d!}~\int_{\M_{g,1}} \kappa_1^{3g-2-d}  \psi_1^{d}~.
%\end{eqnarray}
%This is the volume of $\M_{g,1}$ with respect to  $\omega_{k,\Delta}$, which differs from the usual Weil-Petersson volume of $\M_{g,1}$ computed just with $\kappa_1$.  
We have reduced the counting of black hole microstates to the computation of the intersection numbers $\int_{\M_{g,1}} \kappa_1^{3g-2-d}  \psi_1^{d}$ on moduli space.  

Equation \rref{approx} is just a semi-classical approximation of the number of states. In order to determine the subleading corrections we would need to construct the full quantum Hilbert space.  One natural approach is geometric quantization.
 The first step in this procedure is pre-quantization, where one demands that the symplectic structure %$ {{k\over 2\pi^2}  \omega_{WP} + \Delta \psi_1} $, viewed as a  cohomology class, %is an element of integral cohomology in $H^2(\M_{g,1},\Z)$.  This allows us to view $\omega$ as the 
is the first Chern class of a line bundle ${\cal L}_{k,\Delta}$:
\be
c_1(\cL_{k,\Delta}) = k \kappa_1 + \Delta \psi_1~.
\ee  
%In this picture, $\omega_{k,\Delta}$ is viewed as the first Chern class of a line bundle $\cL_{k,\Delta}$ with 
In this picture, the quantization of $k$ and $\Delta$  follow from the existence of a pre-quantum line bundle.  
Indeed, the line bundle associated with $\Delta$ is just the canonical line bundle describe above.
Similarly, $\kappa_1$ is the first Chern class of a line bundle (see e.g. \cite{MR815769, takhtajan1991}).  
%The result is that for any integer $(k,\Delta)$ one must construct a line bundle $\cL_{k,\Delta}$ on $\M_{g,1}$ whose first Chern class is 
%\be
%c_1(\cL_{k,\Delta}) = \omega_{k,\Delta}~.
%\ee  
The wave functions are sections of the line bundle $\cL_{k,\Delta}$.  

The precise spectrum can be determined only once we fix some additional structure on moduli space. Since $\M_{g,1}$ is K{\"a}hler \cite{MR1745010} one natural approach is K{\"a}hler quantization of $\M_{g,1}$. 
The Hilbert space of microstates is then the space $H^0({\M}_{g,1},\cL_{k,\Delta})$ of holomorphic sections of $\cL_{k,\Delta}$ on $\M_{g,1}$.  Each such holomorphic section describes a particular black hole microstate.  
Unfortunately, these sections are difficult to describe precisely.  
The dimension of $H^0({\M}_{g,\cdot},\cL_{k,\Delta})$ can be estimated using the Hierzebruch-Reimann-Roch theorem, which  gives the Euler characteristic for the line bundle $\cL_{k,\Delta}$ % in terms of characteristic classes on $\M_{g,1}$:
\be\label{rr}
\sum_i (-1)^i \dim H^i (\M_{g,1},\cL_{k,\Delta}) = \int_{\M_{g,1}} td(\M_{g,1}) e^{k\kappa_1  +  \Delta \psi_1} 
\ee 
where $td(\M_{g,1})$ is the Todd class of $\M_{g,1}$.  The semi-classical approximation \rref{approx} then amounts 
 to the assumption that the Todd class can be ignored and that the higher cohomology groups $H^i (\M_{g,1},\cL_{k,\Delta})$ vanish for $i>0$.  For the time being we will be content with our semi-classical approximation \rref{approx}, though it would be interesting to investigate the subleading corrections to $N_g(k,\Delta)$ determined by \rref{rr}.
 
\subsection{Two Conjectures on the Asymptotics of Intersection Numbers}

To apply \rref{approx} we must compute the following intersection numbers on the moduli space of Riemann surfaces:
\be
I_{g,d} \equiv \int_{\M_{g,1}} \kappa_1^{3g-2-d}  \psi_1^{d}~.
\ee 
Closed form expressions for $I_{g,d}$ exist only in certain cases.  
Some very simple cases at low genus can be computed explicitly.  More complicated intersection numbers involving $\psi$ classes can be computed using the recursion relations conjectured by Witten  \cite{MR1144529} and proven by Kontsevich \cite{MR1171758}.
For example, intersections of just $\psi_1$ classes were computed by Itzykson and Zuber \cite{MR1180858}
\be\label{psi}
I_{g,3g-2} = {1\over 24^g g!}~.
\ee
More recently, Mirzakhani \cite{MR2257394} presented an different set of recursion relations which allow one to compute any $I_{g,d}$.  This allowed Zograf \cite{zograf} to conjecture a formula, later proven \cite{Mirzakhani:2011gta}, for the asymptotic behaviour of these intersection numbers:
\be\label{zog}
I_{g,d} \approx (2g)! {(3g-2-d)!\over d!} \alpha^g \beta^d g^\gamma \delta^d + \dots
\ee
for large $g$ at fixed $d$.  We are interested in the factorial behaviour, but the constants $\alpha, \beta, \gamma, \delta$ that control the subleading dependence on $g$ and $d$ are known \cite{Mirzakhani:2011gta}.  In fact, we expect this formula to hold even at large $d$, provided that we take $d \ll g$. 

We are interested in the slightly more general case where $g$ and $d$ are large but may be of the same order.  We will make conjectures for these numbers in two separate limits, both of which interpolate between the known results \rref{psi} and \rref{zog}. % with $g$ and $d$ and $P(g,d)$ is an unknown function which scales as at most a power of $g$ and $d$ in this limit.  This power law scaling will not be important to us.  

Our first conjecture is that at large but fixed $g$ the intersection numbers scale exponentially with $d$, rather than factorially, so that
\be\label{conjecture1}
%\log I_{g,d} \approx -d D_g + E_g + \dots
I_{g,d} \approx A_g B_g^{-d}
\ee
at large $d$ and fixed (large) $g$.   
Here $A_g$ and $B_g$ are positive, genus dependent numbers.  We have exhibited only the leading dependence: subleading terms which are power law in $d$ have been suppressed. 
This exponential dependence  will be enough for us to make our desired statement about the number of states at fixed genus.  However, we can go one step further and determine how the coefficients $A_g$ and $B_g$ scale with $g$ by demanding that they are consistent with \rref{psi} and \rref{zog}.
We find that $A_g$ should increase factorially and $B_g$ should increase quadratically as $g\to\infty$:  
\be\label{scalings}
 A_g\approx (5g)!,~~~~~B_g \approx C g^2~.
\ee 
where $C$ is a genus-independent constant. %\footnote{Comparing with the numerical values in \cite{zograf} we can estimate $C\approx 1.1$.}
We have neglected here all terms which are subleading in the large $g$ limit, such as power law and exponential corrections to $A_g$ and linear corrections to $B_g$. %\footnote{Some subleading terms can be estimated as well.  For example, we expect $B_g \approx (3g-2) - \log (3g-2)$.  At $g=20$ this gives $B_{20}\approx 53.9$, in (surprisingly good) agreement with the numerical result in figure 4.} 

%Similarly, we expect $B$ to scale linearly with $g$, in order to be consistent with \rref{psi} and \rref{zog}.
% in the $\dots$ we have suppressed all of the terms which are subleading in $d$.
%We have chosen to write the factorials in \rref{conjecture} in a way which will be convenient for us below, but we could, for example, write the factorial dependence more simply as ${2g! (3g-d)! / d!}$ by adjusting the values of $A$, $B$ and $P(g,d)$.

The conjecture \rref{conjecture1} should be easy to study given the work of \cite{Mirzakhani:2011gta}; we hope to report on this in the future.  We have compared this conjecture to values of intersection numbers up to genus $20$ supplied by P. Zograf \cite{zograf:private}.  The results at genus 20 are summarized in Figure 4.

  \begin{figure}[htbp]\label{f1}
\centering
\includegraphics[scale=1.1, clip=true]{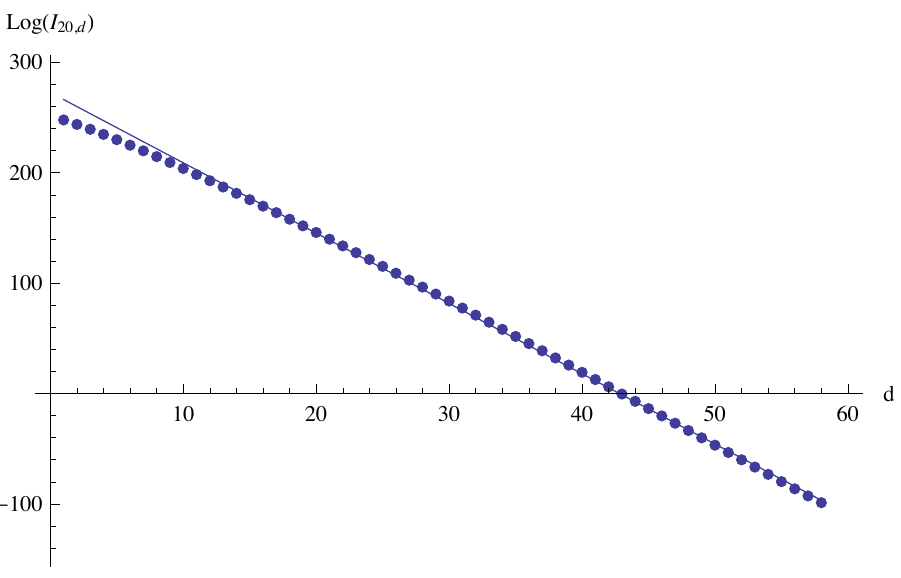}
         \caption{The intersection numbers $I_{20,d}$ at genus 20 exhibit exponential scaling with $d$ for $d \gtrsim 10$.  The linear fit of $\log(I_{20,d})$ (indicated by the blue line) gives coefficients $B_{20}\approx 600$ and $A_{20}\approx 3 \times10^{118}$ in equation \rref{conjecture1}.  Numerical data supplied by P. Zograf.}
\end{figure}

%The above conjecture is sufficient to compute the number of black hole microstates at fixed genus, but we are also be interested in understanding the 
One can also study the scaling with genus by considering intersection numbers in the limit $g\to \infty$ with the ratio 
%Our second conjecture is that when $g$ is taken to be large with the ratio 
$a = d/g$ held fixed.  In this case, the natural conjecture is that
\be\label{conjecture2}
% I_{g,a g} \approx {(5g)! \over (2a g)!}
 I_{g,a g} \approx {\left((5-2a)g\right)!}
\ee
as $g\to \infty$.  
Again, we include here only the factorial dependence; there will be subleading terms which are exponential  in $g$.  This matches with \rref{psi} and \rref{zog} when $a=0$ and $3$.  The $a=3/2$ behaviour is compared to numerical values of the intersection numbers up to genus 20 in Figure 5.  This conjecture is more difficult to verify numerically; in the future we hope to have more numerical data to check this behaviour.

  \begin{figure}[htbp]\label{f1}
\centering
\includegraphics[scale=1.1, clip=true]{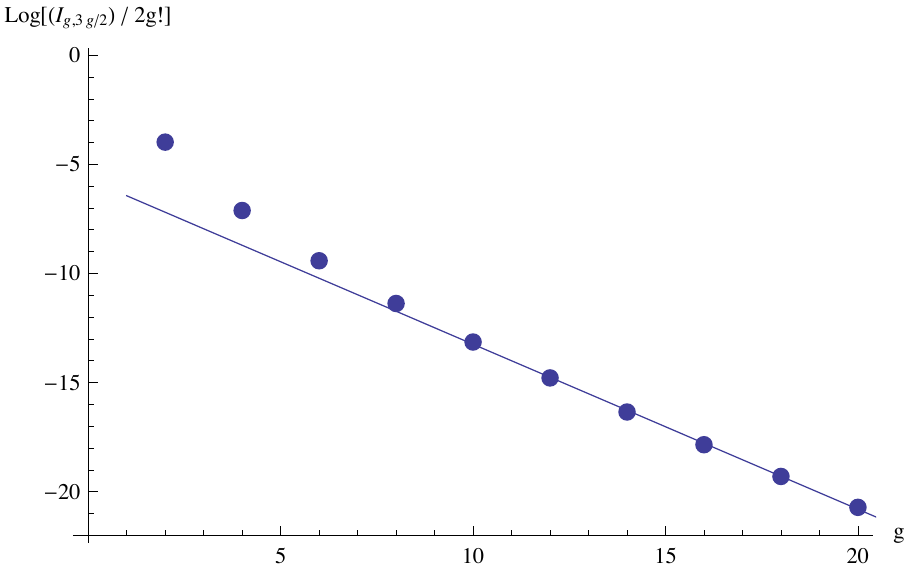}
         \caption{The linear behaviour of $\log(I_{g,3g/2}/ (2g)!)$ (indicated by the blue line) with $g$ indicates that the intersection numbers $I_{3g/2,3g/2}$ scale approximately as $(2g)!$, in accordance with \rref{conjecture2}.  Numerical data  supplied by P. Zograf.}
\end{figure}

\subsection{Counting Microstates at Fixed Genus}

We can now count black hole microstates using \rref{approx}.  We will start by considering microstates with a large but fixed genus $g$ behind the horizon.  If we are interested in the number of states at large $\Delta$, we can use \rref{conjecture1} to find
\bea\label{fixedg}
N_g(k,\Delta) &\approx& A_g 
\sum_{d=0}^{3g-2} { k^{3g-2-d} (\Delta/B_g)^d  \over (3g-2-d)!~ d!}
\cr
&\approx & {A_g\over 3g!} \left(k+{\Delta \over B_g}\right)^{3g-2}~.
\eea
%The good news is that this is finite.  
So if the topology behind the horizon is bounded, the number of quantum states is finite.
For large black holes ($\Delta \gg k$) the number of states $N_g \approx \Delta^{3g-2}$ is far smaller than the semi-classical Bekenstein-Hawking result, $N(\Delta)\approx e^{4\pi \sqrt{k \Delta}}$. 

We conclude that the quantization of geometry at fixed genus does not give enough states to account for the  semi-classical black hole entropy.  
%Only far from the classical regime (when $k\approx {\cal O}(1)$) does this result have a hope of completely accounting for the black hole entropy.  It is therefore tempting to speculate that the only theories of ``pure" quantum gravity in AdS$_3$ -- i.e. theories with only metric degrees of freedom -- are highly quantum mechanical theories with AdS radius of order one in Planck units.\footnote{  One possible piece of evidence in favour of this may come from the AdS/CFT correspondence. A natural criterion, first proposed by Witten \cite{}, that a CFT must obey in order to be regarded as dual to a pure theory of quantum gravity is that it is {\it extremal} in the sense that all primary states have dimension $\Delta > k$ (Witten in addition required that the CFTs be holomorphically factorized, but it is not know whether this condition is necessary).  In these theories all primary states in the theory can be regarded as black holes, rather than perturbative degrees of freedom.  The only known examples of CFTs with this property are the Ising model ($c={1/ 2}$), the tri-critical Ising model ($c={7/ 10}$) and the Monster CFT ($c=24$).  In each case dual theories of pure gravity have been proposed \cite{}.  The fact that all of these theories have $c= {k/ 24} \approx {\cal O} (1)$ may reflect the fact that the quantization of geometry does not correctly account for black hole entropy in the semi-classical limit.}

\subsection{An Explosion at Large Genus}

%To count the total number of black hole microstates with area $L$, we must sum over all possible topologies behind the horizon.  

%In order to obtain a large entropy, it is necessary to consider very large genus behind the horizon.  
To obtain a large number of microstates, it appears that we must consider the limit of large genus behind the horizon.  
Indeed, one might expect that the vast number of microstates of a large black can be understood only when one considers a ``quantum foam" with very complex topology behind the horizon.  
%It is therefore tempting to associate the entropy \rref{sum} with a ``quantum foam" behind the horizon, with many degrees of freedom arising from the increasing topological complexity behind the horizon. 
However, it is difficult to make this notion precise in the present context.  From the results of the previous section (e.g. \rref{scalings} and \rref{fixedg}) the number of states contains terms that scale like $N_g \approx (2g)!$ at large genus.
%\be
%N_g\approx {2g!} \left(k+{\Delta \over C g^2}\right)^{3g}~.
%\ee
The total number of states should, at least in principle, include a sum over genera.  This sum diverges badly, even at small $\Delta$.  

We are therefore faced with a tension: the fixed genus entropies are too small, but the large genus contributions are naively divergent and vastly outnumber the Bekenstein-Hawking entropy.  It is interesting to speculate on how this tension might be resolved.  We can not, for example, simply put an upper bound on the genus: the resulting entropy will not increase linearly in the area of the black hole.  One might, for example, decide to only include terms in  \rref{approx} with $d\ge a g$ for some fixed $a$.  From \rref{conjecture2} we see that the sum over genus appears to converge if we take $a>1$.\footnote{Taking the sum over genus with fixed $a$ gives contributions to the entropy that scale as a power $\Delta$.  It does not, however, reproduce the Bekenstein-Hawking entropy for any $a$.}

One simple possibility is that the semi-classical counting formulas fail at large genus and we must understand the properties of the Hilbert space in terms of, for example, the space of holomorphic sections $H^0({\M}_{g,1},\cL_{k,\Delta})$.  %The sum over genera might thus be rendered finite.
Another possibility is chiral gravity differs in some subtle way from the naive quantization of geometry described here.  It may be that one must impose further conditions on the space of states in quantum gravity.  Naively distinct classical configurations -- such as those with different genus $g$ -- may be secretly related by some new type of gauge symmetry.  Speculations on this, albeit in somewhat different settings, appear in \cite{noscoop, Castro:2011zq}.  

\subsection{An Entropy Proportional to Horizon Area?}

We conclude by discussing a feature of this computation which suggests how one might obtain an entropy linear in the horizon area, albeit with a coefficient too small to account for the semi-classical Bekenstein-Hawking entropy.  

The full sum over genus gives a number of states which can be written as
\bea
N(k,\Delta) = \sum_{g=0}^\infty N_g(k,\Delta) %&=& \sum_{g=0}^\infty \sum_{d=0}^{3g-2} k^{3g-d-2} \Delta^d {I_{g,d}\over (3g-2-d)! d!} \cr
&\approx&\sum_{d=0}^\infty {1\over d!} \left({\Delta\over k}\right)^d \left(\sum_{g\ge {d+2\over 3}} {I_{g,d} \over (3g-2-d)!} k^{3g-2}\right)
\eea
where we have reordered the sum over genus and the sum over $d$. 
Let us now focus on those terms with $g \gg d$, where we can apply \rref{zog} to find contributions of the form 
\bea
N(k,\Delta) \approx  \sum_{d=0}^\infty {1\over (d!)^2 } \left(\beta {\Delta\over k}\right)^d \left(\sum_{g\gg d} (2g)!~ \alpha^g~ k^{3g-2} + \dots \right) 
\eea
Keeping only the leading term in the sum over $g$, the sum over $d$ has a saddle point with
\be\label{crazy}
N(k,\Delta) \approx e^{\sqrt{\beta {\Delta / k}}}  \left(\sum_{g\gg d} (2g)!~ \alpha^g~ k^{3g-2} + \dots \right)
\ee
The leading term in the sum over $g$ in \rref{crazy} is infinite but independent of $\Delta$.

Our speculative proposal is that % the sum over $g$ diverges for the same reason that 
we should  treat %Borel resum 
the sum over genus in \rref{crazy} just as we would any other divergent loop expansion.  In particular, we should view it as an asymptotic series whose divergence signals the presence of non-perturbative effects of order $e^{k^{-3/2}}$. We will then proceed by assuming that these non-perturbative effects render the sum finite. % non-perturbative effects. 
% associated with this resummation.  

The result of this manipulation is 
\be\label{sis}
S(k, \Delta) =\log N({k,\Delta}) \approx \sqrt{\beta} L + \dots
\ee
This entropy is proportional to the area of the event horizon!  
It is, however, given by the area in AdS units, not Planck units.
We conclude that only far from the classical regime (when $k\approx {\cal O}(1)$) does this method have a hope of completely accounting for black hole entropy.  %Perhaps the procedure described above is the correct one, and indicates that the only ``pure" theories of quantum gravity in AdS$_3$ -- i.e. theories with only metric degrees of freedom -- are highly quantum mechanical theories with AdS radius of order one in Planck units.

%An important feature of the entropy formula \rref{sum} is that, in order to obtain an entropy that grows linearly with horizon area it is necessary that the genus $g$ behind the horizon is at least of order $\Delta/k\sim L$.   
%
%This observation, however, brings up an elephant in the room which we have so far done our best to ignore: at large genus, the number of states diverges rapidly due to the $C_g\approx (2g)! k^{3g-2}$ prefactor in \rref{sum}.  %At fixed genus, the entropy is linear in area only for black holes with $\Delta/k \lesssim {\cal O}(g)$.  Thus to achieve this scaling for arbitrarily black holes it is necessary to include geometries with high genus.  

%A final possibility is that new degrees of freedom -- in addition to those of gravity -- must be included in the theory.
%We hope to comment more definitively on this in the future.

\

\textbf{Acknowledgments:} I thank T. Blais, N. Do, J. Kim, D. Marolf, R. Mazzeo, M. Mirzakhani, M. Porrati, S. Ross, J. Teschner, R. Vakil, E. Verlinde, H. Verlinde, E. Witten and P. Zograf for useful discussions.  I am especially grateful to P. Zograf for supplying numerical data on large genus intersection numbers \cite{zograf:private} and to J. Kim and M. Porrati for sharing a preliminary copy of their paper \cite{noscoop}.    
This research is supported by the National Science and Engineering Research Council of Canada.

\footnotesize

\bibliographystyle{utphys}

\bibliography{bibliography}

\end{document}